\PassOptionsToPackage{dvipsnames}{xcolor}
\documentclass{article}
\usepackage{spconf,amsmath,graphicx}
\usepackage{times}
\usepackage{latexsym}
\usepackage[utf8]{inputenc}
\usepackage{microtype}
\usepackage{soul}
\usepackage{url}
\usepackage{color}
\usepackage[dvipsnames]{xcolor}
\usepackage{multirow}
\usepackage{graphicx}
\usepackage{amsmath}
\usepackage{amsthm}
\usepackage{booktabs}
\usepackage{algorithm}
\usepackage{algorithmic}
\usepackage{CJKutf8}
\usepackage{amsfonts}
\usepackage{amssymb}
\usepackage{bbm}

\title{IMPROVING THE MODALITY REPRESENTATION WITH MULTI-VIEW CONTRASTIVE LEARNING FOR MULTIMODAL SENTIMENT ANALYSIS}
\name{Peipei Liu$^{1,2}$, Xin Zheng$^{2,3}$, Hong Li$^{1,2*}$\thanks{Corresponding Author: Hong Li lihong@iie.ac.cn}, Jie Liu$^{1,2}$, Yimo Ren$^{1,2}$, Hongsong Zhu$^{1,2}$, Limin Sun$^{1,2}$ }
\address{$^{1}$ School of Cyber Security, University of Chinese Academy of Sciences \\ $^{2}$ Institute of Information Engineering, Chinese Academy of Sciences \\ $^{3}$School of Software, Henan University}
\begin{document}
%
\maketitle
\begin{abstract}
  Modality representation learning is an important problem for multimodal sentiment analysis (MSA), since the highly distinguishable representations can contribute to improving the analysis effect. Previous works of MSA have usually focused on internal fusion strategies for different modalities within one sample, and the external usage of cross reference relations among different samples was given less attention. Recently, the rise of contrastive learning provides powerful clues for us to learn modal representation with stronger discriminative ability. In this study, we explore the approach of representations improvement and devise a three-stages framework with multi-view contrastive learning to refine representations for the specific objectives. Firstly, for each modality, we employ the supervised contrastive learning to pull samples within the same class together while the other samples are pushed apart. Then, a self-supervised contrastive learning is designed for the distilled cross-modal representations after a novel Transformer-based interaction module. At last, we leverage again the supervised contrastive learning to enhance the fused multimodal representation.We conduct extensive experiments on three open datasets, and results show the advance of our model.
\end{abstract}

\section{Introduction}
\label{sec:intro}

Sentiment analysis, which aims to identify the emotional state of people based on specific information, is a significant task on natural language processing (NLP) as it can be applied for opinion mining, dialogue generation, and user behavior analysis \cite{CHSIMS}. Early work mainly focused on textual sentiment analysis and achieved fruitful results \cite{Pang2007O}. However, the expressions of human emotion are usually diverse and only text cannot be sufficient to understand human behaviors and intents. Compared with the text content, multimodal data including spoken words, vision and acoustic can provide richer information. For instance, facial attributes and the tone of voice in videos can alleviate the ambiguity of textual information through expressive emotional tendencies\cite{SunMM22}. For this reason, the research on multimodal sentiment analysis is attracting more and more attention.

Existing study of MSA focuses on efficient multimodal fusion methods based on the common feature extraction approaches like Convolutional Neural Network (CNN), Long Short-term Memory (LSTM) network and Transformer. These fusion methods \cite{MSACat, AVVA} can be divided into two categories by their manner: early-fusion and late-fusion. The former is achieved by fusing the features from different modalities immediately after they are extracted, and the fusion feature is further used for predicting the final outputs\cite{SunMM22, MultimodalTrans}. On the contrast, in late-fusion, multiple modalities are trained end-to-end independently, and then the prediction results from different modalities are used for voting the final decision\cite{LMF,TFN}. 
Despite the success, they have not worked better on the study that how to learn similar representations for semantically similar content and build more discriminative feature representation with the external usage of cross reference relations among different samples, which is often sufficient to distinguish the target from other objects.

Recent works of the contrastive learning (CL) enlighten us on improving such representation ability. The core idea of the contrastive learning is pulling together an anchor and “positive” samples in embedding space while pushing apart the anchor from “negative” samples, and it can fall into self-supervised contrastive learning (SSCL) \cite{SCCL, BingoICLR, Consert, HeCVPR} and supervised contrastive learning (SCL)\cite{supercl_nips}. The SSCL constructs “positive” pairs through the anchor and its co-occurrences or augments while “negative” pairs by randomly choosing samples matching with the anchor. SCL is an extension of SSCL, the only difference is that “positive” pairs are from the same class but not the same original sample.

In this paper, we mainly explore the improvement method for modality representations of MSA by creating a three-stages framework with multi-view contrastive learning. Specifically, after extracting the initial multimodal features (i.e., vision, text, and audio), we first employ the supervised contrastive learning to independently cluster and learn unimodal hidden representations of each sample since the label information is available. As a result, samples belonging to the same unimodal class have the higher semantic similarity. Then, we design and introduce the self-supervised contrastive learning based on frozen unimodal representations at the second stage. In this stage, we propose a multi-level cross-modal Transformer module for refining specific modality-aimed cross-modal representations. After that, the refined representations of each sample from different attention-views are applied to build the pairs for self-supervised contrastive learning. At the third stage, we fuse the resulting representations from the second stage into the multimodal representations by self-attention, and carry out again supervised contrastive learning to get the most discriminative multimodal representation of each sample for the final classification task.

Our main contributions can be summarized as follows:
\begin{itemize}
\item[1)]  We propose a novel framework with multi-view contrastive learning for producing the more discriminative representation to improve the performance of MSA task. To the best of our knowledge, this is the first work of contrastive learning used for MSA task.   
\item[2)]  We extend the cross-modality Transformer to adopt a cross-attention mechanism within each level at multi-level fusing more modalities to obtain the most effective multimodal information.
\item[3)]  Extensive experiments are conducted on the open datasets, and the results demonstrate the advance of our method. We also give the careful experimental analysis to understand the model better and further seek the possible future directions.
\end{itemize}

\section{method}
\label{sec:pagestyle}
\begin{figure*}[h]
  \centering
  \includegraphics[height=2.4in,width=5.6in]{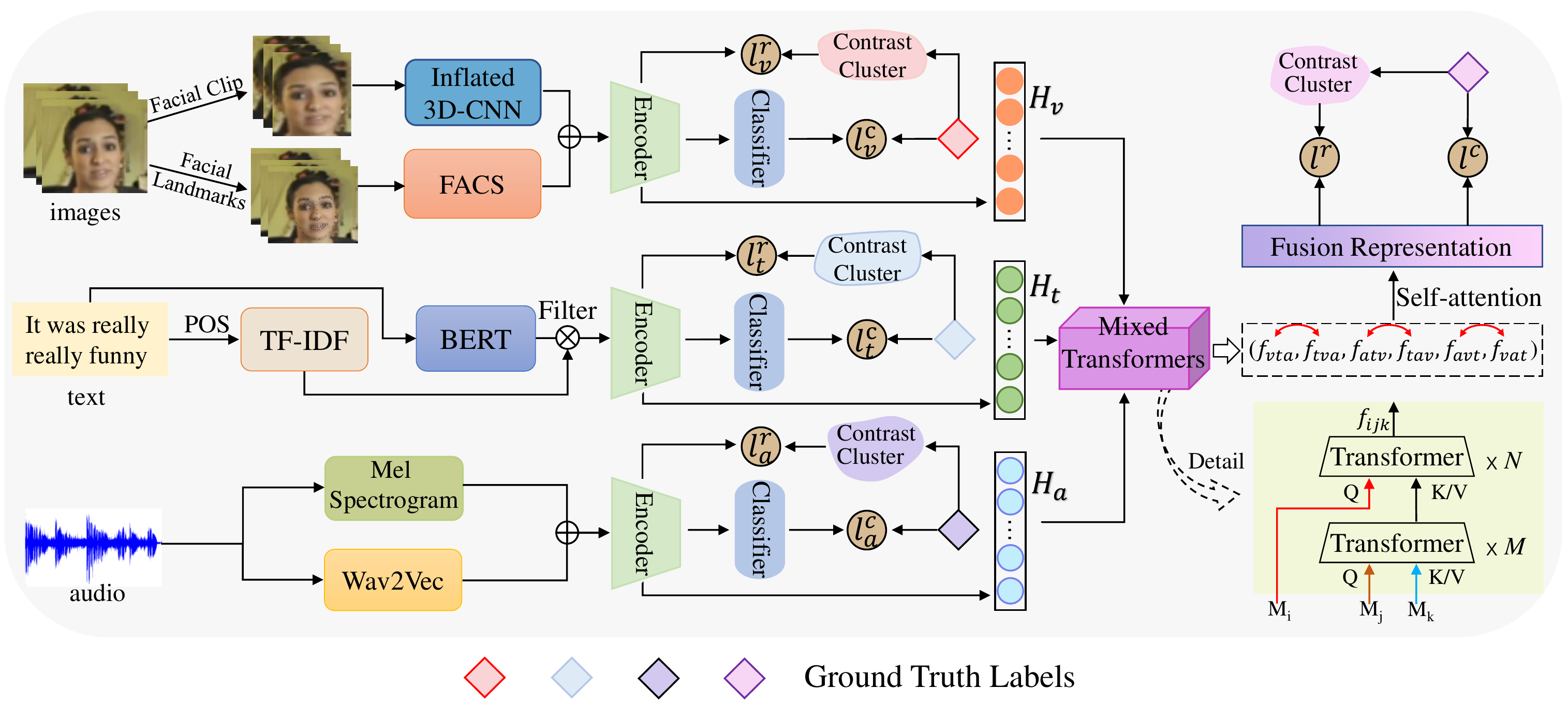}
  \caption{The overview of our proposed method.}  
  \label{model_overview}
  \vspace*{-0.4cm}
\end{figure*}

In this section, we first briefly explain the definition of MSA, and then give the details of our model. The overview of the model is show in Figure.\ref{model_overview}.

\textbf{Task definition.} The task of MSA aims to predict sentiment intensity, polarity, or emotion label with considering three modalities \textit{t} (text), \textit{v} (video) and \textit{a} (acoustic). Usually, a MSA dataset with \textbf{N} samples can be regard as \{($t_1$, $a_1$, $v_1$, $y_1$), ..., ($t_N$, $a_N$, $v_N$, $y_N$)\}, where $y_i$ is the ground truth label of \textit{i}-th sample. Moreover, for the modality-specific multi-task MSA which extra annotates each modality, we can formulate the dataset as \{($t_1$, $a_1$, $v_1$, $y_{t,1}$, $y_{a,1}$, $y_{v,1}$, $y_1$),...,($t_N$, $a_N$, $v_N$, $y_{t,N}$, $y_{a,N}$, $y_{v,N}$, $y_N$)\}, where ($y_{t,i}$, $y_{a,i}$, $y_{v,i}$, $y_i$) is the label for \textit{i}-th sample and $y_{t,i}$, $y_{a,i}$, $y_{v,i}$ are the labels for \textit{t}, \textit{a}, \textit{v} respectively.

\subsection{Feature Extraction}
Before the training of our MSA task, we apply tools and pre-trained models to extract the initial feature for each modality.

\textbf{\textit{Video}}
In order to obtain richer spatial and temporal information in video, we have adopted two extraction methods for the feature. On the one hand, we use MTCNN \footnote{\url{https://github.com/ipazc/mtcnn}} to align the faces in the video, and use the Inflated 3D-CNN \cite{I3DCNN} model to extract the video stream feature. On the other hand, we use the visual feature extraction method adopted in \cite{CHSIMS}, using the MTCNN algorithm to align faces, and then using OpenFace2.0 \cite{SENA} to extract frame-level FACS features. We concatenate the both feature for the next encoding.

\textbf{\textit{Text}}
The pre-trained BERT-base word embeddings are employed to obtain word vectors for the text. It is necessary to note that we do not use all the words for textual feature but only a part of the sequence. Considering that only nouns, verbs, adjectives and adverbs are important for the sentiment, we first filter the words with these Part-of-Speech\footnote{The tool we use is Spacy: \url{https://spacy.io/}} and then use TF-IDF to select the words with the maximum value.

\textbf{\textit{Acoustic}}
Similar to the extraction of visual features, we also use two models to extract speech features. We use the toolkit LibROSA and wav2vec2.0 \cite{SENA} to extract audio features from audio sequences. Also, we concatenate the features for next use.

\subsection{Contrastive Training}
For the convenience in following introduction, we treat $X^i_m\in\mathbb{R}^{L\times d}$ as the extracted feature of $i$-th sample, where $m\in\{v, a, t\}$ and $L$ represents the length of sequence and $d$ denotes the feature dimension.

\textbf{\textit{SCL for Unimodal Representation (SCL-1)}}

As the above description, we train the model with labels to learn representation making the similar samples closer than the other different in the embedding space.

First, we take the extracted feature $X^i_m$ of each modality as input the unimodal encoder to get hidden representation $H^i_m$, and then a projection network maps the $H^i_m$ to a vector $z^i_m$ used for predicting whether two samples are from the same class.  
{\setlength{\abovedisplayskip}{2pt}
\setlength{\belowdisplayskip}{2pt}
\begin{eqnarray}
  H^i_m = Encoder(X^i_m)
\end{eqnarray}
\begin{eqnarray}
  z^i_m=Projection(H^i_m)
\end{eqnarray}}
where $Encoder(\cdot)$ is the Transformer and $Projection(\cdot)$ is a multi-layer perceptron (MLP) with the output dimension 256.

Given a batch with $N$ samples, where the anchor $i\in I$=$\{1, ..., N\}$, a contrastive loss can be defined as follows:
{\setlength{\abovedisplayskip}{1pt}
\setlength{\belowdisplayskip}{1pt}
\begin{footnotesize}
\begin{eqnarray}
  loss^r_m=\sum_{i \in I}{-\frac{1}{|P^{(i)}_m|} \sum_{p \in P^{(i)}_m}log{ 
    \frac{exp(sim(z^i_m, z^p_m)/\tau)}{\sum_{a \in A^{(i)}_m}exp(sim(z^i_m, z^a_m)/\tau)}
  }}
\end{eqnarray}
\end{footnotesize}}
where $m\in\{v, a, t\}$, $r$ represents \textit{representation}. For the modality $m$, $P^{(i)}_m$ is the positive set whose samples are from the same class with $i$, $A^{(i)}_m$ is the set of all samples in $I$ except for $i$ (i.e., $A^{(i)}_m=I-\{i\}$). $sim(\cdot,\cdot)$ denotes cosine similarity between two vectors and $\tau$ is a temperature parameter.

We train $loss^r_t$, $loss^r_a$, $loss^r_v$ separately. After the contrastive training, the $Encoder(\cdot)$ is used for downstream classification task and $Projection(\cdot)$ is discarded. We will investigate whether finetuning the parameters of $Encoder(\cdot)$ in experiments. 

\textbf{\textit{SSCL for Refined Modality Representation (SSCL)}}

After obtaining the representation $H^i_m$ of each modality $m$ for $i$-th sample, we design a multi-level cross-modal Transformer module to refine specific modality-aimed cross-modal representations according to the other two modalities. Concretely, for the input $(Q,K,V)$ of Transformer, we first have $H^i_v=Q$ and $H^i_t=K/V$ to derive the vision-aware text representation $f^i_{vt}$. Then, we assign $H^i_a$ to $Q$, relatively pure $f^i_{vt}$ to $K$ and $V$ to compute the most effective simultaneously vision- and acoustic- aware text representation $f^i_{avt}$. We can also get the $f^i_{vat}$, $f^i_{tva}$, $f^i_{vta}$, $f^i_{tav}$, $f^i_{atv}$ by this way.

An intuitive opinion to the two representations which are from different refinement orders but for the same modality is that they should have the less gap. Based on this view, we create the self-supervised contrastive learning for each sample. Previous works of self-supervised contrastive learning construct the positive pairs by each anchor and its augment, but we here pair the two different representations of one sample such as ($f^i_{avt}$, $f^i_{vat}$). For the $t$ modality, the batch will result in $2N$ representations $\{f^1_{avt}, f^1_{vat}, ..., f^N_{avt}, f^N_{vat}\}$ and each data point is trained to find out its counterpart among 2N-1 other samples. As the same in \textbf{SCL-1}, we need to map $\{f^1_{avt}, f^1_{vat}, ..., f^N_{avt}, f^N_{vat}\}$ to $\{z^{1}_{t}, z^{1}_{t'}, ...,z^{N}_{t}, z^{N}_{t'}\} $. The contrastive loss  $loss^d_t$ can thus be defined as:
{
  \setlength{\abovedisplayskip}{3pt}
  \setlength{\belowdisplayskip}{3pt}
  \begin{scriptsize}
    \begin{equation}
    \begin{split}
    -(\sum_{i \in I}{
      log \frac{exp(sim(z^i_{t}, z^i_{t'})/\tau)}{\sum_{k \in I}exp(sim(z^i_{t}, z^k_{t'})/\tau) + \sum_{j \in I, i\neq j}exp(sim(z^i_{t}, z^j_{t})/\tau)}}+\\
      \sum_{i \in I}{
      log \frac{exp(sim(z^i_{t'}, z^i_{t})/\tau)}{\sum_{k \in I}exp(sim(z^i_{t'}, z^k_{t})/\tau) + \sum_{j \in I, i\neq j}exp(sim(z^i_{t'}, z^j_{t'})/\tau)}})
     \end{split}
    \end{equation}
  \end{scriptsize}
  }
where the $sim(\cdot, \cdot)$ and $\tau$ are the same as above. We train the loss of all three modalities jointly, and the self-supervised contrastive loss in this section can be expressed by:
  \begin{eqnarray}
    loss^d = loss^d_t+loss^d_a+loss^d_v 
  \end{eqnarray}

After the contrastive training, we can get the learned representation ($f^i_{avt}$, $f^i_{vat}$, $f^i_{tva}$, $f^i_{vta}$, $f^i_{tav}$, $f^i_{atv}$). Next, we use the linear function $L(\cdot)$ and the self-attention $SA(\cdot)$ to compute the final fusion representation $f^i$. That is:
  \begin{eqnarray}
    f^i = SA(L(f^i_{avt}), ..., L(f^i_{atv}))
  \end{eqnarray}

\textbf{\textit{SCL for Multimodal Representation (SCL-2)}}

We try again the supervised contrastive learning in this section. The learning process is the same with \textbf{SCL-1}. The only difference is that we operate on the multimodal representation instead of the specific modality, i.e., we replace $H^i_m$ with $f^i$ in Eq.(2). The contrastive loss for the fusion representation after projecting $f^i$ to $z_i$ is:
{
\begin{footnotesize}
\begin{eqnarray}
  loss^r=\sum_{i \in I}{-\frac{1}{|P^{(i)}|} \sum_{p \in P^{(i)}}log{ 
    \frac{exp(sim(z^i, z^p)/\tau)}{\sum_{a \in A^{(i)}}exp(sim(z^i, z^a)/\tau)}
  }}
\end{eqnarray}
\end{footnotesize}}

\subsection{Classifier Training}
For both the modality-specific multi-task MSA and multimodal MSA, we train the MLPs as the output layer. For the unimodal sentiment analysis, we use the learned $H^i_m$ for training whereas for multimodal MSA, we use the learned $f^i$. Following \cite{MISA}, we use the standard cross-entropy loss for classification tasks while mean squared error loss for the regression tasks. The following shows the multimodal task:
{\setlength{\abovedisplayskip}{2.5pt}
\setlength{\belowdisplayskip}{1.5pt}
\begin{eqnarray}
  y'_i = W_o(ReLU(W_f(f^i)+b_f))+b_o
\end{eqnarray}
\begin{eqnarray}
  loss_c^c = -\frac{1}{N}\sum_{i\in I}{y_ilog{(y'_i)}}
\end{eqnarray}
\begin{eqnarray}
  loss_r^c = -\frac{1}{N}\sum_{i\in I}{||y_i-y'_i||^2_2}
\end{eqnarray}}

\section{Experiments}
\label{sec:typestyle}
\subsection{Datasets}
We conduct experiments on three datasets, CMU-MOSI\cite{zadeh2016mosi}, CMU-MOSEI \cite{zadeh2018multimodal} and CH-SIMS\cite{CHSIMS}, respectively. The MOSI dataset contains 93 videos from YouTube segmented into 2199 video clips. The MOSEI dataset collects 23453 video clips from 1000 speakers and 250 subjects.The CH-SIMS dataset collects 60 videos from movies, TV series and variety shows, and divides these videos into 2281 video clips. Each video clip in CH-SIMS is annotated with three independent single-modal labels and one multi-modal label. Both MOSI and MOSEI datasets are regression tasks, and CH-SIMS includes regression task and classification task. For the regression task, we treat the same score as the same class for contrastive learning.

\subsection{Parameters Setting}
For video features, we use the ResNet-50 I3D model trained on the Kinetics-400 dataset to extract frame-level features in 2048 dimensions, in which 32 frames of pictures are collected for each video segment, and the collection step size is 2. We set the fps in OpenFace to 30, and obtain facial landmarks in 3D features, rigid face shape and Non-rigid face shape features, head pose features, gaze related features, action units features, a total of 709 dimensions of frame-level facial features.
For audio features, the LibROSA method is used to extract 33-dimensional frame-level audio features. The specific configuration parameters are consistent with \cite{CHSIMS}. The method of wav2vec2.0 is used to extract 768-dimensional frame-level audio features. The batch-size N is set as 128, and the hidden dimension of Transformer is 512. The temperature parameter $\tau$ is set as 0.2.

\subsection{Metrics}
Similar to previous work, we evaluate the model's f1 score (F1) and 2-class accuracy(Acc), while mean absolute error (MAE) and Pearson correlation (Corr) are also used as evaluation metrics. Higher values of F1 score, Acc and Corr represent better performance, while lower values of MAE are better.

\subsection{Baseline Models}
We choose several excellent Early Fusion methods and Late Fusion methods as our baseline models.\\\textbf{TFN} \cite{TFN} learns both the intra-modality and inter-modality dynamics through multiple subnetworks and a new multimodal fusion approach respectively.\\\textbf{LMF} \cite{LMF} designs a low-rank multimodal fusion method to decompose the weight tensor and accelerate fusion process.\\\textbf{MulT} \cite{MultimodalTrans} proposes the multimodal transformer, where a cross-modal attention module is learned for reinforcing one modality's features with those from the other modalities. \\\textbf{MISA} \cite{MISA} factorizes each modal information into modality-invariant and modality-specific to learn common features across modalities and reduce the modality gap. \\\textbf{CHSIMS} \cite{CHSIMS} designs a multi-modal multi-task learning framework, which not only performs the multimodal prediction, but also predicts the single-modal results.

\subsection{Experimental Results}
The Table.\ref{main} reports the results of our model compared with others on the dataset MOSI and MOSEI. We can find that, in most metrics, our model can achieve the best results. We think that the contrastive learning is effective on the representation feature augmentation. The Acc-2 metric of our model on MOSI is lower than the MulT, that maybe the TF-IDF in our model gets the insufficient valid words and affects the overall model. Also, the MAE of our model is higher than MISA on MOSEI, that because MISA has learned the better modality-invariant feature. In addition to these two datasets, we also conduct the experiments on CHSIMS to check the performance of our model on modality-specific prediction task.

\begin{table}[]
  \scriptsize
  \begin{tabular}{ccccccccc}
  \hline
  Model & \multicolumn{4}{c}{MOSI}      & \multicolumn{4}{c}{MOSEI}     \\ \cline{2-9} 
        & MAE   & Corr  & Acc-2 & F1    & MAE   & Corr  & Acc-2 & F1    \\ \hline
  TFN   & 0.901 & 0.698 & 80.8  & 80.7  & 0.593 & 0.700 & 82.5  & 82.1  \\
  LMF   & 0.917 & 0.695 & 82.5  & 82.4  & 0.623 & 0.677 & 82.0  & 82.1  \\
  MulT  & 0.861 & 0.711 & \textbf{84.1}  & 83.9  & 0.58  & 0.703 & 82.5  & 82.3  \\
  MISA  & 0.804 & 0.764 & 82.1  & 82.03 & \textbf{0.568} & 0.724 & 84.23 & 83.97 \\ \hline
  Ours  & \textbf{0.769} & \textbf{0.783} & 83.7  & \textbf{84.2}  & 0.573 & \textbf{0.741} & \textbf{84.95} & \textbf{85.01} \\ \hline
  \end{tabular}
\caption{The Result Comparison on MOSI and MOSEI}
  \label{main}
  \end{table}

\begin{table}[]
  \scriptsize
  \centering
  \begin{tabular}{cccccc}
  \hline
  Task               & Model  & Acc-2          & F1             & MAE            & Corr           \\ \hline
  \multirow{2}{*}{A} & CHSIMS & 67.70          & 79.61          & 53.80          & 10.01          \\
                     & Ours   & \textbf{68.25} & \textbf{80.13} & \textbf{52.71} & \textbf{13.24} \\ \hline
  \multirow{2}{*}{V} & CHSIMS & 81.62          & 82.73          & 49.57          & 57.61          \\
                     & Ours   & \textbf{83.76} & \textbf{84.19} & \textbf{47.93} & \textbf{58.04} \\ \hline
  \multirow{2}{*}{T} & CHSIMS & 80.26          & \textbf{82.93} & \textbf{41.79} &  49.27            \\
                     & Ours   & \textbf{81.47} & 81.32          & 42.07          & \textbf{49.33}       \\ \hline
  \end{tabular}
  \caption{The Results of our model and CHSIMS on the CHSIMS dataset}
  \label{resuni}
  \vspace{-0.3cm}
  \end{table}
  From the Table.\ref{resuni}, we also find that our model can achieve the satisfactory effect on the modality-specific task. In fact, we train the unimodal classifier on the frozen representation learned by supervised contrastive learning on the single-modality task. But when we train the final multimodal classifier, we finetune the unimodal representation $H_m$ and the refined representation ($f^i_{avt}$, $f^i_{vat}$, $f^i_{tva}$, $f^i_{vta}$, $f^i_{tav}$, $f^i_{atv}$). 
  In the future work, we will try the self-supervised contrastive learning on the same modality using several different feature extractors since they actually aim to the same task. For example, we can contrast the acoustic features from LibROSA and wav2vec. Also, we may explore the other applications by using the contrastive learning.

\section{Conclusion}
In this paper, we propose a novel framework with multi-view contrastive learning for improving the modality representation used for the multimodal sentiment analysis. And we also conduct the extensive experiments and experimental analysis to demonstrate the advance of our method.

\vfill\pagebreak

\bibliographystyle{IEEEbib}
\bibliography{strings,refs}

\end{document}